\documentclass[]{basi}
\usepackage{amsmath,amssymb,pstricks,wasysym,stmaryrd,enumerate,epsfig}
\usepackage{graphics,fancyhdr,natbib,psfig}
\usepackage{floatflt,graphicx}
\usepackage[T1]{fontenc}
\usepackage{times}

\newcommand{\dg}{$^{\circ}$}

\begin{document}
\title{The Galactic centre region}
\author[Subhashis Roy]
{Subhashis Roy\\
National Centre for Radio Astrophysics (TIFR), \\
Pune University Campus, Post Bag No.3, Ganeshkhind, Pune 411
007, India
\thanks {e-mail: roy@ncra.tifr.res.in}}
\date{Received 9 August 2004; accepted 16 September 2004}
\maketitle

\begin{abstract}
We have observed the central 45$^\prime$ region of the Galaxy at 620 MHz band of the
Giant Metrewave Radio Telescope (GMRT) in radio continuum, and measured the 
polarisation properties of 64 small
diameter background extragalactic sources seen through the $-$6\dg$<l<$6\dg,
$-$2\dg$<b<$2\dg\ region with the Australia Telescope Compact Array (ATCA) and the 
Very Large Array (VLA).  Our 620 MHz observations
show that Sgr~A* is located behind the HII region Sgr~A~West. Using the
ATCA and the VLA observations, we measured the Faraday rotation measure (RM) of
the polarised sources. The measured RMs are mostly positive, and show no
reversal of sign across the rotation axis of the Galaxy.  This rules out any
circularly symmetric model of magnetic field in the region.  We estimate the
magnetic field strength in the region to be $\sim 10\mu$G, which raises
doubts against an all pervasive mG field in the central few hundred pc
of the Galaxy.
\end{abstract}

                                                                                                           
\section{Introduction}
Being located two orders of magnitude closer than the nearest large galaxy,
the Galactic centre (GC) can be studied with much higher spatial resolution
and sensitivity than is possible for other galaxies. Because of this
advantage, we can identify unique objects like the Radio-arc consisting of
linear parallel filaments (Yusef-Zadeh et al. 1984), or the 2.6$ \times 10^6$
M$_{\odot}$ black hole most probably associated with the compact radio
source Sgr~A*  (Ghez et al. 1998).  These studies have also established a
concentration of gas, mostly in the form of molecular clouds characterised by
densities $\sim$10$^4$ molecules cm$^{- 3}$, velocity dispersions $\sim$10 km
s$^{- 1}$, temperatures $\sim$70 K and apparent magnetic fields $\sim$1
mG in the region. Most of these quantities are one to two orders of
magnitude larger than those found in the disk of our Galaxy. The review by
Morris \& Serabyn (1996), describes this research. Melia \& Falcke (2001)
provide the recent advances made in the research on the black hole and the
central region (a few pc) of the Galaxy.  The three-dimensional source
distribution in the GC region is complex, and despite the studies mentioned
above, identification of the objects and the physical processes operating in
the region are difficult to understand, some of which we discuss below.

From the high-resolution ($\sim$~arcsec) observations with the 
VLA at radio frequencies (Pedlar et al. 1989), the following
sources within the central 15$^\prime$ of the Galaxy have been identified or
reconfirmed.  (i) At the dynamical centre of the Galaxy is the compact
nonthermal radio source known as Sgr~A*. (ii) Around Sgr~A* is the HII region
Sgr~A~West, whose morphology resembles a face-on spiral galaxy.  (iii) Near
Sgr~A~West is Sgr~A~East, which is believed to be a supernova remnant (SNR).
(iv) A 7$^\prime$ halo, which has been proposed to be a mixture of thermal and
non-thermal emission (Pedlar et al. 1989). 

Sgr~A* (see Melia \& Falcke 2001 for a review) was not detected at
frequencies below 960 MHz and observations at 408 MHz (Davies et al. 1976)
and at 330 MHz (Pedlar et al. 1989) provided upper limits ($\le$0.1 Jy at 330
MHz) on its flux density. Sgr~A* probably has a low frequency turnover below
1 GHz, but the nature of the turnover has never been examined in detail
(Melia \& Falcke 2001). Recently, Nord et al. (2004) claim to have detected
Sgr~A* at 330 MHz. However, their detection at 330 MHz remains provisional
(Roy \& Rao 2004).

Magnetic fields can be strong enough to have a significant influence on the
dynamics and evolution of the sources in the central region of our Galaxy.
Magnetic pressure can contribute significantly to the overall pressure balance
of the interstellar medium (ISM) and can even influence the scale height of
the gas. It can affect the formation and motion of clouds and perhaps mediate
in the star formation process (Beck et al. 1996). 
Therefore, it is important to measure the magnetic field geometry and strength
in the central part of the Galaxy. However, other than the central 200 pc
region, no systematic study has been made to estimate the magnetic fields in
the inner 5 kpc region of the Galaxy (Davidson 1996). The estimates
within the central 200 pc region are mainly based on the observations of the
non-thermal filaments or NTFs (Gray et al.  1995).  However, if the NTFs are
manifestations of a peculiar local environment (Shore \& Larosa 1999),
inferences drawn from these observations can be misleading.

Here we describe (i) the 620 MHz GMRT observations of the 
GC, which was aimed at studying the low frequency spectrum and the location
of Sgr~A* with respect to Sgr A West HII region,
(ii) ATCA and VLA observations of 65 background sources located in the central 
$-$6\dg$<l<$6\dg, $-$2\dg$<b<$2\dg\ region of the Galaxy to measure Faraday rotation
measure (RM), which could yield a model dependent magnetic field in the GC
region.

\section{Observations and data reduction}
\subsection{GMRT observations of the GC at 620 MHz}
The 620 MHz observations of the GC were carried out with the GMRT with a
nominal bandwidth of 16 MHz.  Full-synthesis ($\sim$7 hours) observations were performed.
Absolute flux density calibration was performed using 3C48 and 3C286 data.
The increase of $T_{sys}$ from the calibrator field to the target source
affects the source visibility amplitudes in the default observing mode
(i.e. the Automatic Level Control (ALC) in the system is turned on), and
this was tackled by (i) multiplication of the target source data with ratio of
total power (when ALCs were off) between source and calibrator, or (ii)
observations with ALC off and using unnormalised cross-correlation data (see Appendix A).

The data were processed within Astronomical Image Processing System (AIPS)
using standard programs. Images of the fields were formed by Fourier inversion
and Cleaning (IMAGR). The initial images were improved by phase only
self-calibration. To improve the deconvolution of the extended emission, we
made the final image using Multi-resolution Clean (Wakker \& Schwarz 1988).
The GMRT map at 620 MHz (Fig.~1) has a dynamic range of about 150, and is
limited by systematics.

\subsection{RM observations of the background sources seen towards the GC}
A total of 37 sources were observed with the ATCA using their 4.8 GHz and 8.5
GHz bands. The ATCA observations were made using a 6~km array configuration.
Each target source was typically observed for a total of 40--50 minutes. Since
we used an E-W array configuration, to get a
satisfactory {\it uv}-coverage each source was observed $\approx$10 times
almost equally spaced in hour angle.  Calibration and editing of the ATCA
data were performed using {\sc MIRIAD}. The calibrated visibilities were imaged
using standard {\sc AIPS}.

We used the VLA in its BnA configuration to observe 27 relatively weak
sources in the sample. The default continuum mode, with a single frequency
channel of bandwidth 50 MHz in each IF band was used. Observations were
centred at frequencies of 4.63, 4.88, 8.33 and 8.68 GHz.  Each source was
observed for typically 5 minutes at 2 different hour angles. All the data were
calibrated and processed using the {\sc AIPS} package.

\section{Results and Discussions}
\subsection{GMRT observations of the GC at 620 MHz}
The central 15$^\prime$ region of the Galaxy at 620 MHz is shown in Fig.~1. The
compact source Sgr~A*, along with other well known sources like Sgr A East and
the 7$^\prime$ halo are clearly seen.  To compare the smaller-scale features near
Sgr~A~West with higher frequency maps, we have overlayed in Fig.~2, the 620 MHz
map in gray scale on the 4.8 GHz VLA map in contours convolved to a common
resolution. There is almost one to one correspondence between the brighter
emission features at 4.8 GHz comprising the Sgr~A~West region and a drop in the
total intensity at 620 MHz (indicated by a white region seen along Sgr A West),
indicating that the thermal emission is optically thick near 620 MHz.

\begin{figure}
\begin{minipage}[t]{0.62\textwidth}
\includegraphics[clip=,angle=0,width=\textwidth]{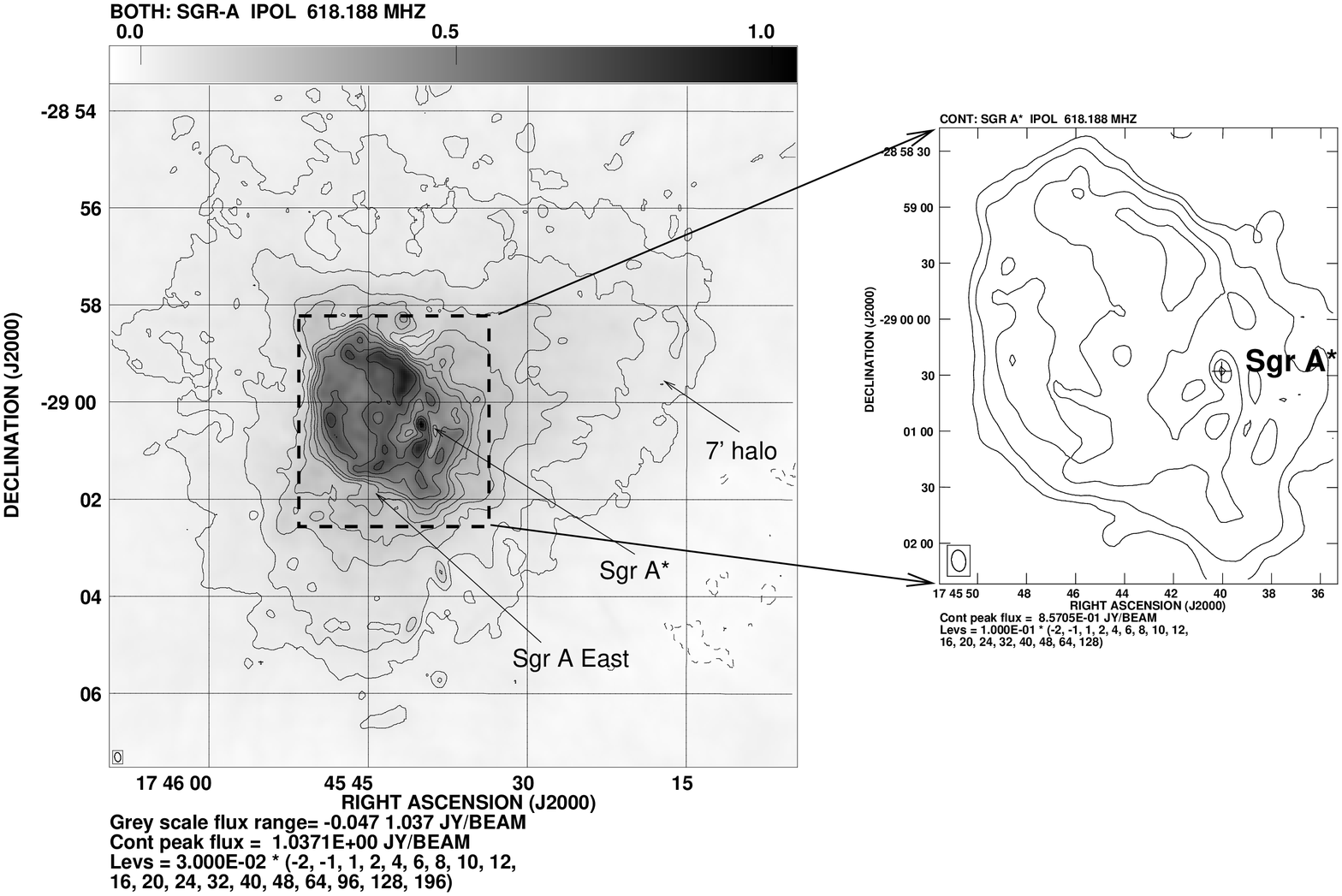}
\caption{620 MHz GMRT map of the central 15$^\prime$ region of the Galaxy
(left). The rms noise is 6.5 mJy/beam, and the beamwidth is
11.$^{\prime\prime}$4 $\times$ 7.$^{\prime\prime}$6.  The central part of the
image with the known position of Sgr~A* marked is shown in the right.}
\end{minipage}
\hfill
\begin{minipage}[t]{0.33\textwidth}
\includegraphics[clip=,angle=0,width=\textwidth]{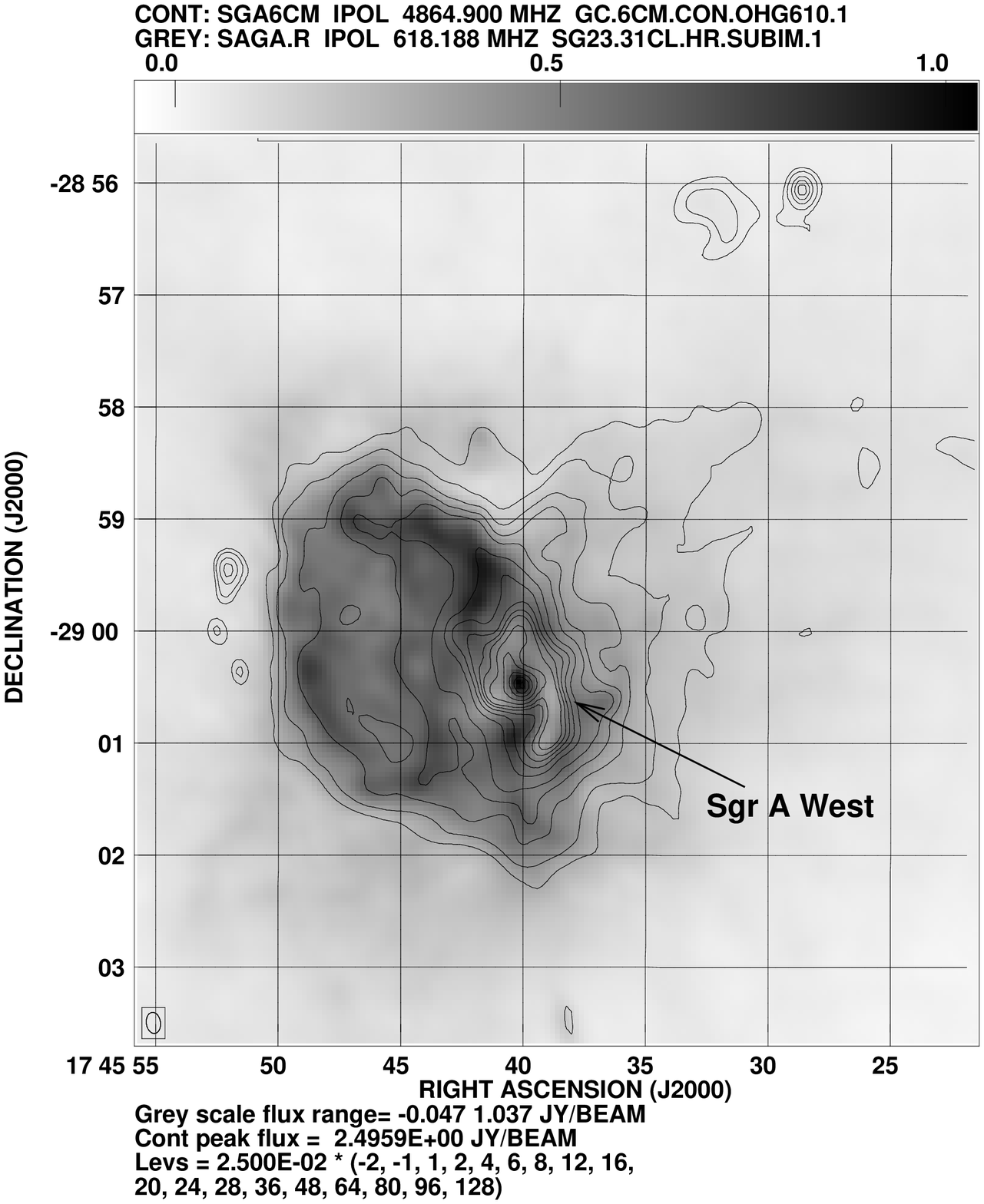}
\caption{The 4.8 GHz continuum map of the Sgr A complex (Yusef-Zadeh 1989) in
contours overlaid on the 620 MHz gray scale map of the same region. Resolution
is 11.$^{\prime\prime}$4 $\times$ 7.$^{\prime\prime}$6.}
\label{620mhz.gc.central.map}
\end{minipage}
\end{figure}

\subsubsection{Flux density of Sgr~A*}
\label{sgra*.flux}
While estimating the flux density of Sgr~A* at 620 MHz, we reduced the
confusion due to the extended emission by imaging only those visibilities
having {\it uv} distance $>$7 k$\lambda$. The flux density estimated from the
image plane is about 0.5 $\pm$0.1 Jy. We note that even after applying a lower
{\it uv} cutoff, there is significant background confusion within a beam of
size 7.$^{\prime\prime}5 \times 4^{\prime\prime}$. This confusion causes an uncertainty of about
0.1 Jy in the estimated flux density.

\subsubsection{Low frequency spectral index of the Sgr~A*}
Although the high radio frequency spectrum of Sgr~A* is well established, the
spectrum below 1.4 GHz is not well estimated. At 1.4 GHz, the flux density of
Sgr~A* is about 0.5 Jy and its time averaged spectral index, $\alpha$ defined as
S$\propto\nu^{\alpha}$, between 1.4 and
8.5 GHz is +0.17 (Melia \& Falcke 2001). Davies et al. (1976) found the flux
density of Sgr~A* at 960 MHz to be a factor of two less than at 1.6 GHz and
suggested that it has a low frequency turnover around 1 GHz. This appears to
be confirmed from their upper limit of 50 mJy at 408 MHz and the 100 mJy
upper limit set by Pedlar et al. (1989) at 330 MHz. However, the present 620
MHz observations along with its tentative 330 MHz detection (Nord et al.
2004) raise questions about earlier measurements.  Based on the average flux
density of Sgr~A* at 1.4 GHz (Zhao et al. 2001) and its known spectrum
between 1.4 and 8.5 GHz, we expect a flux density of 0.44 Jy at 620 MHz,
which is consistent with our measurements within the error-bars.
Therefore, at frequencies of a few GHz to 620 MHz, the observed spectral
index of Sgr~A* is nearly flat.

Though the advection-dominated accretion flow (ADAF) model of emission from
Sgr~A* fails to explain the observed low frequency emission, ADAF along with
self-absorbed synchrotron emission either from a relativistic jet, or from a
small relativistic fraction of electrons embedded in the accretion flow
(Yuan et al. 2003) could explain the observed spectrum of Sgr~A* from X-ray
ranges to low radio frequencies of about 620 MHz.

\subsection{Location of Sgr~A*}
At 620 MHz, Sgr~A~West shows evidence for free-free self-absorption.  In Roy
\& Rao (2004), we have shown the optical depth of Sgr A West HII region near
Sgr~A* to be 2.5$\pm$0.5.  If Sgr~A* was located behind Sgr~A~West, then its
flux density would have been attenuated by a factor of 10 due to thermal
absorption by Sgr A West.  However, the spectral index of Sgr~A* between 620
MHz and 1.4 GHz is roughly the same as between 1.4 GHz and 8.5 GHz showing no
effect of free-free absorption by Sgr~A~West. This indicates that Sgr~A* is
located in front of Sgr~A~West. It is possible to invoke alternate scenarios
like holes in Sgr~A~West along the line of sight to Sgr~A*, but these appear
to be unlikely (see Roy \& Rao 2004 for details).

\subsection{Magnetic field in the GC: RM observations of the background
sources}
Based on the ATCA data, 24 sources were found to have at least one polarised
component. The FWHM sizes of the beams in the ATCA 4.8 GHz images are
$\approx$6$^{\prime\prime} \times 2^{\prime\prime}$, and the typical rms noise is 0.23 mJy/beam in
Stokes I and about 0.15 mJy/beam in Stokes Q and U. Of the sources observed
using the VLA, 21 have at least one polarised component. The VLA images have a
beam of FWHM size $\sim$2$^{\prime\prime} \times 1.^{\prime\prime}5$ and the rms noise is
typically 75 $\mu$Jy/beam. A map of one of the sources is presented in Fig.~3.

\begin{figure}
\begin{minipage}[t]{0.45\textwidth}
\includegraphics[clip=,angle=0,width=\textwidth]{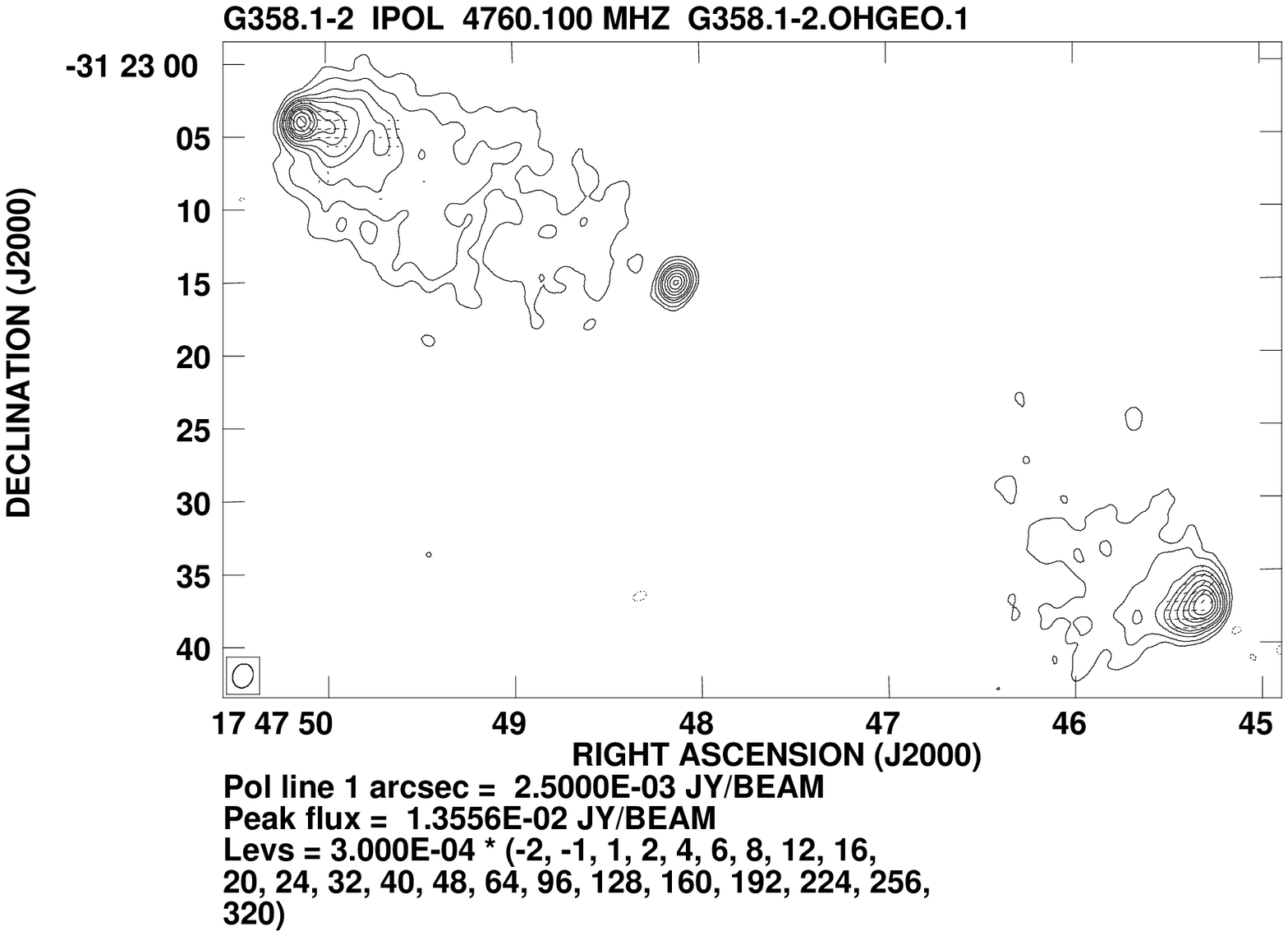}
\caption{4.8 GHz map of one of the sources (G358.1-2) observed in the RM
study.}
\end{minipage}
\hfil
\begin{minipage}[t]{0.45\textwidth}
\vspace{-5.0 cm}
\includegraphics[clip=,angle=270,width=\textwidth]{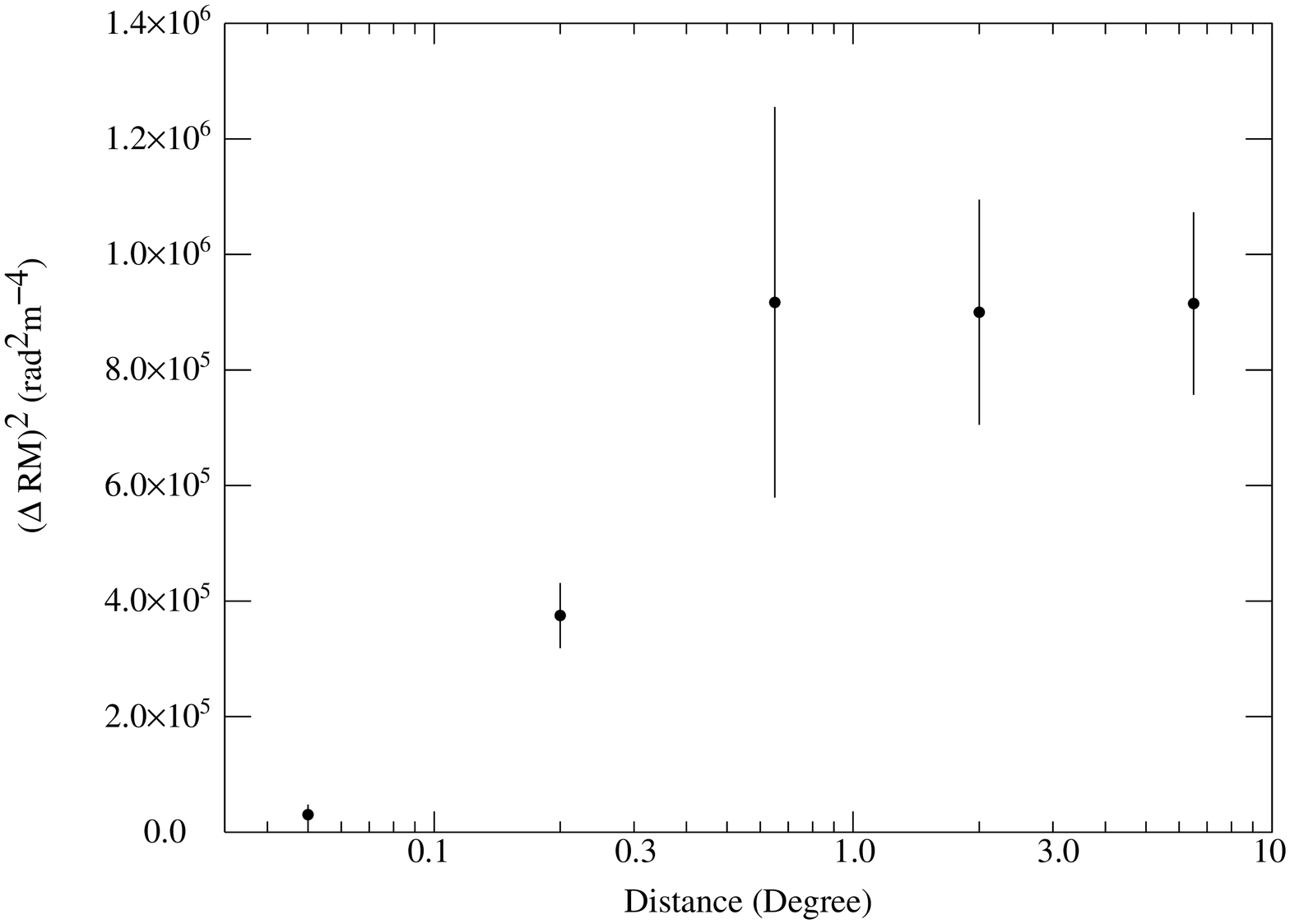}
\caption{Plot of the structure function of the RM.}
\end{minipage}
\end{figure}

\subsubsection{Features in the RM sky near the GC}
The estimated RM towards 45 sources are plotted in Fig.~5.  In the plot, the
positive RMs have been indicated by cross ($\times$) and the negative RMs by
a circle ($\bigcirc$), where the size of the symbols increases linearly with the 
$|$RM$|$.  The region appears to be dominated by positive RM.
The median value of the RMs is estimated to be 467$\pm$87 rad m$^{- 2}$.

To detect small-scale fluctuations in the RM, we have plotted the RM structure
function in Fig.~4. As can be seen from the plots, the RM structure function
tends to zero as the separation of the polarised components tends to zero. Since
intrinsic RMs of different sources are independent of their angular separations, 
our result supports an interstellar origin (plasma turbulence) of the RM towards the GC.
The structure function tends to saturate for components separated by more than
0.\dg3. From the plot, the coherence scale (where the structure function
attains half of its maximum value) of the Faraday screen is estimated to be
about 0.\dg2\ (30 pc at a distance of 8.5 kpc).

\begin{figure}
\centering
\hbox{
\includegraphics[clip=,angle=270,width=0.9\textwidth]{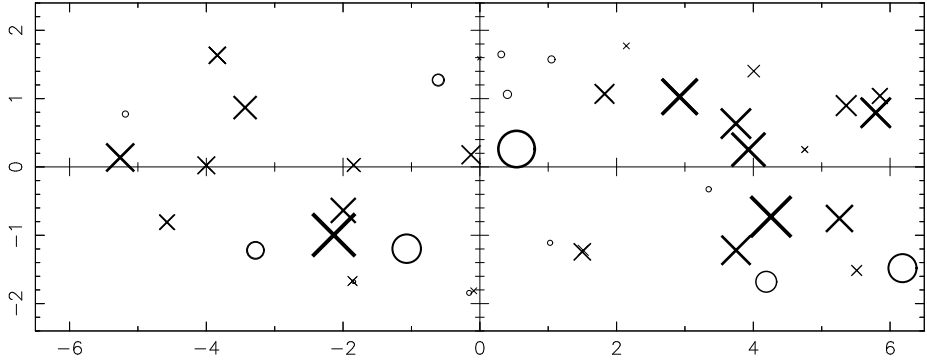}
}
\caption{The plot of the RM towards various sources as a function of the Galactic
longitude and latitude. The cross symbol ($\times$) indicates positive RM and the circle
($\bigcirc$) negative RMs, where the symbol size increases linearly with $|$RM$|$.}
\label{gc.rm}
\end{figure}

\subsubsection{Geometry of the magnetic field along the Galactic plane} 
Our observations show a large-scale positive bias in RM towards the GC, with a
mean RM of 525$\pm$129 rad m$^{- 2}$ towards the sources with positive Galactic
longitude, and  389$\pm$103 rad m$^{- 2}$ towards the sources with negative
Galactic longitude.  The absence of any reversal of sign of the RM (4$\sigma$
result) across the rotation axis of the Galaxy ($l$=0\dg), indicates that the
magnetic field across the rotation axis of our Galaxy is not circularly
symmetric.  Two models, the axisymmetric and the bisymmetric spiral model of
the magnetic fields are generally invoked to explain the field geometry in
galaxies (Beck et al. 1996).  It is thought that current loop generated by
rotation of the matter around the centre (dynamo model) gives rise to the
axisymmetric spiral configuration of the magnetic field. However, if the
magnetic fields at the present time are the result of concentration of a
primordial magnetic field due to collapse of matter while the Galaxy was
formed, then it gives rise to the bisymmetric spiral configuration of the
magnetic field in galaxies. In our Galaxy, the magnetic field configuration
has been suspected to be bisymmetric (Simard-Normandin \& Kronberg 1980), in
which case, the field does not undergo a reversal of sign across the centre.
Our observations support the bisymmetric spiral configuration in the inner
parts of our Galaxy.

\subsubsection{Estimation of the magnetic field from the RM}
In order to estimate the line of sight ({\it los}) component of the magnetic fields from the
RM, we need a model of the electron density in the region.  Based on the 
Very Long Baseline Array (VLBA)
observations of the scattering diameter of extragalactic sources
(Bower et al. 2001), we consider an electron density of 0.4 cm$^{-3}$ in the
region, which is twice the value predicted by the Taylor \& Cordes (1993)
model for the inner Galaxy. If the central 2~kpc is considered responsible
for the Faraday rotation, then the {\it los} averaged magnetic field is
estimated to be 0.7 $\mu$G (Roy, Rao \& Subrahmanyan 2004).

To compute the random component of the magnetic field, we note that the typical
scale size of the Faraday screen as estimated in $\S$3.3.1 is $\sim$30 pc.
Hence, there will be about 70 cells of irregularities within the central 2~kpc
of the Galaxy. Assuming that the RM contributions from the individual cells
along the {\it los} are random about the mean, the cell contributions would
have an rms RM value of 83 rad m$^{-2}$.  If the rms electron density within
such a cell is 0.4 cm$^{-3}$, the estimated magnetic field is about 9 $\mu$G,
which is more than an order of magnitude higher than its {\it los} average
value. This small field raises doubts about an all-pervasive mG field in the
central few hundred pc of the Galaxy (Yusef-Zadeh \& Morris 1987). 
The large-scale picture suggested by our observations is  that the
magnetic field even a few degrees away from the GC is oriented nearly
perpendicular to the {\it los} pointing towards us on both sides of the GC.

\section*{Acknowledgements}
GMRT is run by the National Centre for Radio Astrophysics of the Tata Institute
of Fundamental Research.  The Australia Telescope is funded by the Commonwealth
of Australia for operation as a National Facility managed by CSIRO. The
National Radio Astronomy Observatory is a facility of the National Science
Foundation operated under cooperative agreement by Associated Universities,
Inc.


\section*{References}
Beck, R., Brandenburg, A., Moss, D., Shukurov, A., Sokoloff, D. 1996, ARA\&A, 34, 155\\
Bower, G. C., Backer, D. C., Sramek, R. A. 2001, ApJ, 558, 127\\
Davidson, J. A. 1996, in ASP Conf. Ser. 97: Polarimetry of the Interstellar Medium, 504\\
Davies, R. D., Walsh, D., Booth, R. S. 1976, MNRAS, 177, 319\\
Ghez, A. M., Klein, B. L., Morris, M., Becklin, E. E. 1998, ApJ, 509, 678\\
Gray, A. D., Nicholls, J., Ekers, R. D., Cram, L. E. 1995, ApJ, 448, 164+\\
Melia, F., Falcke, H. 2001, ARA\&A, 39, 309\\
Morris, M.,  Serabyn, E. 1996, ARA\&A, 34, 645\\
Nord, M. E., Lazio, T. J. W., Kassim, N. E., Goss, W. M., Duric, N. 2004, ApJL, 601, L51\\
Pedlar, A., Anantharamaiah, K. R., Ekers, R. D., et al. 1989, ApJ, 342, 769\\
Roy, S.,  Rao, A. P. 2004, MNRAS, 349, L25\\
Roy S., Rao A.P.,  Subrahmanyan R. 2004, submitted to MNRAS \\
Shore, S. N.,  Larosa, T. N. 1999, ApJ, 521, 587\\
Simard-Normandin, M.,  Kronberg, P. P. 1980, ApJ, 242, 74\\
Taylor, J. H., Cordes, J. M. 1993, ApJ, 411, 674\\
Wakker, B. P., Schwarz, U. J. 1988, A\&A, 200, 312\\
Yuan, F., Quataert, E., Narayan, R. 2003, ApJ, 598, 301\\
Yusef-Zadeh, F. 1989, in IAU Symp. 136: The Center of the Galaxy, eds M. Morris, Dordrecht:Kluwere, 243\\
Yusef-Zadeh, F., Morris, M., 1987, AJ, 94, 1178 \\
Yusef-Zadeh, F., Morris, M., Chance, D. 1984, Nature, 310, 557\\
Zhao, J., Bower, G. C., Goss, W. M. 2001, ApJL, 547, L29

\section{Appendices}
\subsection{Appendix A: Keeping the visibility amplitudes constant when the
system temperature ($T_{sys}$) changes}
To perform GMRT observations such that $T_{sys}$ variations have
minimal effect on the observed visibilities, the following procedures can be
adopted.

The Automatic Level Control (ALC) is a negative feedback device that keeps the
output power of the antennas at the optimal levels.
When $T_{sys}$ changes, the ALC changes the effective amplitude gain of the
amplifiers in order to counteract the changes in output power, leading to
changes in the visibility amplitudes.  Therefore, switching off the ALCs avoids
the problem.  However, with ALCs off, there can be sufficient output power to
saturate the electronics system.  Therefore, system Gains need to be adjusted
in a way to avoid this problem.  This is usually achieved by changing the
amplitude gains of all the antennas using attenuators, so that the output power
of all the antennas are equal to their default values. Since, $T_{sys}$ is
maximum when the sky temperature is maximum, this needs to be performed while
observing the strongest source.  Keeping visibility amplitudes free from above
problem also requires data, where the cross-correlation coefficients between
different antenna pairs are not normalised by the total power of the antennas.
As shown in the Appendix-B, it also avoids any possible change of bandshape due
to change of sky temperature.

\subsection{Appendix B: Variation of spectral bandshape in normalised cross
correlated data due to change in sky temperature}
The shape of the instrumental bandpass (bandpass function) is typically
determined by observing a suitably strong calibrator during observations.
It is known that $T_{sys}$ is the sum of the Electronics temperature
($T_{elec}$) and the antenna temperature ($T_{ant}$), and $T_{ant}$ is directly
proportional to the antenna efficiency ($\eta$). Therefore, if $\eta$ is a
function of frequency ($\nu$), then both $T_{ant}$ and $T_{sys}$ are functions
of $\nu$. As normalised visibilities are directly proportional to $\eta$ and
inversely proportional to $T_{sys}$, it implies that the bandpass function is
proportional to $\frac{T_{ant} ( \nu )}{T_{elec} + T_{ant} ( \nu )}$. Since the
bandpass calibration is performed with a calibrator typically away from the
Galactic plane, the $T_{ant} ( \nu )$ may be small and neglected in the
denominator. However, $T_{ant} ( \nu )$ is substantial while observing any
source in the Galactic plane at low radio frequencies ($\lesssim$1.4 GHz).  As
the relationship of the bandpass function with $T_{ant} ( \nu )$ is non-linear,
it changes the overall shape of the bandpass function.

For example, consider the antenna efficiency ($\eta$) to fall from 0.6 at the
centre of the observing band to 0.4 at the edge of the observing band.
Assuming $T_{elec}$=70 K, and the sky temperature ($T_{sky}$) (near Galactic
plane) to be 200 K, the normalised bandpass function at the centre of the band
is equal to $\frac{\eta T_{sky}}{T_{elec} + \eta T_{sky}} = 0.63$, and is
$0.53$ at the edge of the band. However, while observing a bandpass calibrator
away from Galactic plane, the $T_{ant}$ is negligible, ($T_{elec}$ is assumed
to be independent of $\nu$) and the bandpass function is proportional to $\eta$.
However, if this bandshape is used to correct the visibilities of the target
source, the true visibilities at the edge of the observing band will be
overestimated by about 25\%  compared to the centre of the band.

\end{document}